\pdfoutput=1
\documentclass[aps,prl,10pt,twocolumn,floatfix,superscriptaddress,amsmath,amssymb,nofootinbib,longbibliography]{revtex4-2}
\usepackage{graphicx,bm,xcolor}
\usepackage{soul}
\usepackage[colorlinks=true,linkcolor=blue,citecolor=blue,urlcolor=blue]{hyperref}
\newcommand{\ms}{\mu}
\newcommand{\mbh}{M}

\begin{document}\renewcommand{\figurename}{Figure}
\title{Probing Ultralight Dark Matter with Multiband Gravitational Waves: The Full Relativistic Effect of Cloud Self-Gravity}
\affiliation{Department of Physics, College of Sciences, Shanghai University, 99 Shangda Road, 200444 Shanghai, China}
\affiliation{Dipartimento di Fisica, Sapienza Universit\`a di Roma, Piazzale Aldo Moro 5, 00185, Roma, Italy}
\affiliation{INFN, Sezione di Roma, Piazzale Aldo Moro 2, 00185, Roma, Italy}
\author{Zhen Zhong}
\affiliation{Dipartimento di Fisica, Sapienza Universit\`a di Roma, Piazzale Aldo Moro 5, 00185, Roma, Italy}
\affiliation{INFN, Sezione di Roma, Piazzale Aldo Moro 2, 00185, Roma, Italy}
\author{Chen Yuan}
\email[Corresponding author: ]{yuanchen@shu.edu.cn}
\affiliation{Department of Physics, College of Sciences, Shanghai University, 99 Shangda Road, 200444 Shanghai, China}
\author{Xian-Hui Ge}
\email[Corresponding author: ]{gexh@shu.edu.cn}
\affiliation{Department of Physics, College of Sciences, Shanghai University, 99 Shangda Road, 200444 Shanghai, China}
\begin{abstract}
Ultralight boson clouds around rotating black holes emit gravitational waves through annihilation and bound-level transitions. We evolve 33 dominant levels with cloud self-gravity around a black hole of initial mass $10\,M_\odot$ and dimensionless spin $0.9$ at initial couplings $\alpha_0=0.1$ and $0.2$. For an individual source, the dominant gravitational wave channel with the highest signal-to-noise ratio is $322\to211$  at $0.22\,\mathrm{Hz}$ for $\alpha_0=0.1$ and $544\to322$ at $0.90\,\mathrm{Hz}$ for $\alpha_0=0.2$. We also find the $1644\to322$ and $766\to544$ transitions yield the largest source counts at $\alpha_0=0.1$ and $0.2$, respectively. For a four-year observation, the expected counts of Galactic sources at $\alpha_0=0.1$ and $0.2$, respectively, are $0.36$ and $2.8\times10^{-3}$ for Advanced LIGO, $83$ and $2.1$ for Cosmic Explorer, $68$ and $1.2$ for the Einstein Telescope, $3.5\times10^6$ and $2.5\times10^6$ for DECIGO, $1.6$ and $0.017$ for LISA, $5.5$ and $0.094$ for Taiji, and $1.3\times10^2$ and $1.0\times10^1$ for TianQin. Cloud self-gravity strongly suppresses the $211$ annihilation signal in ground-based detectors while making additional decihertz transition lines observable, highlighting the value of multiband gravitational wave observations for probing ultralight dark matter in future detection. 

\end{abstract}
\maketitle

\noindent{\bf{\em Introduction.--}}
The detection of the first gravitational wave (GW) event GW150914 marked the beginning of GW astronomy~\cite{LIGOScientific:2016aoc}. Subsequent observations, from GWTC-1 to GWTC-5, have reported a growing population of merging black holes (BHs)~\cite{LIGOScientific:2018mvr,LIGOScientific:2020ibl,KAGRA:2021vkt,LIGOScientific:2025slb,LIGOScientific:2026wfs}. These systems provide laboratories for testing gravity in the strong-field regime~\cite{Berti:2015itd,Barack:2018yly,Cardoso:2019rvt}. GW observations also probe fundamental fields with extremely weak couplings to ordinary matter~\cite{Barack:2018yly,Brito:2015oca}. Ultralight bosons, as one of the promising dark matter candidates, connect BH dynamics to dark matter through their gravitational interaction~\cite{Hui:2021tkt}. This interaction alone can produce an observable GW signal from a boson cloud around a spinning BH~\cite{Brito:2021war,Riles:2022wwz}.

A rotating BH amplifies a bosonic mode with angular frequency $0<\omega_R<m\Omega_{\rm H}$, where $m$ is the azimuthal number and $\Omega_{\rm H}$ is the horizon angular velocity~\cite{Press:1972zz,Brito:2015oca}. Such an amplification leads to superradiant instability and yields a bosonic cloud around BH, also known as gravitational atom~\cite{Detweiler:1980uk,Brito:2015oca}. The resulting gravitational atom emits GWs through boson annihilation and transitions between energy levels~\cite{Arvanitaki:2010sy,Arvanitaki:2014wva}. Around stellar-mass BHs, the annihilation signal can fall within the frequency band of ground-based GW detectors and the population of stellar-mass BHs contributes a stochastic GW background~\cite{Brito:2017wnc,Brito:2017zvb,Yuan:2021ebu,Yuan:2022bem,Guo:2026yiz}.

For a gravitational atom, multiple levels can grow and radiate GWs, and the BH can reabsorb part of the cloud~\cite{Ficarra:2018rfu,Guo:2022mpr}. Foster and Rodd~\cite{Foster:2026nuz} and Bo\v{s}kovi\'c \textit{et al.}~\cite{Boskovic:2026boh} computed gravitational level transfer through horizon absorption and scalar emission.
Studies of gravitational backreaction have examined synchronized hairy BHs and their stability~\cite{Herdeiro:2014goa,Ganchev:2017uuo,Degollado:2018ypf}, nonlinear cloud evolution~\cite{East:2017ovw,East:2018glu}, and self-gravity corrections to cloud spectra~\cite{May:2024npn,Kim:2025wwj,Takahashi:2026ruw,Boskovic:2026eth,Guo:2026qfb}. Here we compute the complete relativistic fourth-order growth coefficient of $|322\rangle$ on the synchronized family and its GW signatures.

\noindent{\bf{\em Framework.--}}
We consider a complex scalar field of mass $\ms$ with a quadratic potential, minimally coupled to gravity around a Kerr BH of mass $\mbh$ and dimensionless spin $\chi=a/\mbh$. In units $G=c=\hbar=1$, the gravitational coupling is $\alpha=\mbh\ms$. The Klein--Gordon equation on Kerr admits quasi-bound states $|n\ell m\rangle$ with complex eigenfrequencies $\omega=\omega_R+i\omega_I$, where $n$ is the principal quantum number and $\ell,m$ are the angular and azimuthal quantum numbers. These states are unstable, $\omega_I>0$, when they satisfy the superradiant condition $\omega_R<m\Omega_{\rm H}$, where $\Omega_{\rm H}$ is the horizon angular velocity~\cite{Detweiler:1980uk,Brito:2015oca}. The occupation number $N$, which counts the bosons in a state, then grows as $\dot N=2\omega_I N$.

The superradiant instability transfers energy and angular momentum from the BH to the $|211\rangle$ cloud until the superradiant condition is saturated, $\omega_{R,211}=\Omega_{\rm H}$~\cite{Arvanitaki:2010sy,Brito:2015oca}. The BH and the cloud then form a stationary solution, a Kerr BH with synchronized scalar hair~\cite{Herdeiro:2014goa,May:2024npn}. At fixed $\ms$ and Christodoulou mass $\mbh$, these solutions form a family labeled by the occupation number $N_{211}$ of the cloud. We use this family as the self-consistent background. For the systems considered here, we estimate that the cloud rest mass $\ms N_{211}$ reaches only a few percent of the BH mass at saturation. We therefore expand the synchronized solution in the cloud amplitude $\epsilon$ about the Kerr BH at the $|211\rangle$ superradiant threshold:
 \begin{align}
 \Phi&=e^{-i\omega_{R,211}t+i\varphi}
       [\epsilon\Phi^{(1)}+\epsilon^3\Phi^{(3)}+\cdots],\\
 g_{ab}&=g^{(0)}_{ab}+\epsilon^2h^{(2)}_{ab}
                         +\epsilon^4h^{(4)}_{ab}+\cdots ,
 \end{align}
where $g^{(0)}_{ab}$ is the Kerr metric and $\ms N_{211}/\mbh=O(\epsilon^2)$, since the occupation number is quadratic in the field. At leading order, the cloud is a stationary scalar cloud on Kerr~\cite{Herdeiro:2014goa}: $(\Box_{g^{(0)}}-\ms^2)\,\bigl(e^{-i\omega^{(0)}_{R,211}t+i\varphi}\,\Phi^{(1)}\bigr)=0$, with real frequency $\omega^{(0)}_{R,211}$ equal to the horizon angular velocity of the background metric $g^{(0)}_{ab}$. At second order, the cloud's stress-energy tensor sources $h^{(2)}_{ab}$. The metric correction $\epsilon^2h^{(2)}_{ab}$ acts on $\epsilon\Phi^{(1)}$, and the Klein--Gordon equation at $O(\epsilon^3)$ then determines both the profile correction $\Phi^{(3)}$ and the frequency shift in $\omega_{R,211}=\omega^{(0)}_{R,211}+\epsilon^2\delta\omega^{(2)}_{R,211}+O(\epsilon^4)$.

We now ask how the synchronized $|211\rangle$ cloud changes the growth rate of the $|322\rangle$ mode. At $\epsilon=0$, this mode is a scalar perturbation of Kerr, $\delta\Phi\propto\eta\,e^{-i\omega_{322}^{(0)}t+2i\varphi}$, with infinitesimal amplitude $\eta$. For nonzero $\epsilon$, the perturbed stress-energy tensor $\delta T_{ab}\sim\Phi^*\delta\Phi+{\rm c.c.}$, which oscillates at the frequency difference of the cloud and the mode, sources a metric perturbation $\delta g_{ab}=O(\eta\epsilon)$. The perturbation at $O(\eta)$ is then no longer a pure scalar mode, and we expand its scalar and metric parts in powers of $\epsilon$:
\begin{equation}
 \begin{aligned}
 \delta\Phi&=\eta\,[\delta\Phi^{(0)}+\epsilon^2\delta\Phi^{(2)}+\cdots],\\
 \delta g_{ab}&=\eta\,[\epsilon\,\delta g^{(1)}_{ab}+\epsilon^3\delta g^{(3)}_{ab}+\cdots],\\
 \omega_{322}&=\omega^{(0)}_{322}+\epsilon^2\delta\omega^{(2)}_{322}+\epsilon^4\delta\omega^{(4)}_{322}+\cdots,
 \end{aligned}
 \label{eq:pert}
\end{equation}
where $\delta\Phi^{(0)}\propto e^{-i\omega_{322}t+2i\varphi}$. We expand the coupled Einstein--Klein--Gordon equations order by order in $\epsilon$. In Eq.~\eqref{eq:pert}, $\delta g_{ab}$ carries odd powers of $\epsilon$ and $\delta\Phi$ even powers, so each order gives either an Einstein or a Klein--Gordon equation of the same form,
\begin{equation}
 \begin{aligned}
 \mathcal E^{(0)}\bigl[\delta g^{(n)}_{ab}\bigr]&=S^{(n)}_{ab},&\qquad &n\ \text{odd},\\
 \mathcal O^{(0)}\,\delta\Phi^{(n)}&=S^{(n)},&\qquad &n\ \text{even},
 \end{aligned}
 \label{eq:orders}
\end{equation}
where $\mathcal E^{(0)}$ and $\mathcal O^{(0)}$ are the linearized Einstein and Klein--Gordon operators on $g^{(0)}_{ab}$, and the sources contain only lower orders and, for even $n$, the frequency shift $\delta\omega^{(n)}_{322}$. At $O(\eta\epsilon)$, this gives $\delta g^{(1)}_{ab}$. At $O(\eta\epsilon^2)$, $\delta g^{(1)}_{ab}$ acting on the cloud $\Phi^{(1)}$ and $h^{(2)}_{ab}$ acting on $\delta\Phi^{(0)}$ give $\delta\Phi^{(2)}$ and $\delta\omega^{(2)}_{322}$. At $O(\eta\epsilon^3)$, it gives $\delta g^{(3)}_{ab}$. At $O(\eta\epsilon^4)$, it gives $\delta\Phi^{(4)}$ and $\delta\omega^{(4)}_{322}$. We expand the occupation growth rate $2\omega_{I,322}$ in $\ms N_{211}/\mbh$ instead of $\epsilon$:
\begin{equation}
 2\omega_{I,322}=2\omega^{(0)}_{I,322}+\gamma_1\frac{\ms N_{211}}{\mbh}+\gamma_2\Bigl(\frac{\ms N_{211}}{\mbh}\Bigr)^2+\cdots,
 \label{eq:bnet}
\end{equation}
where $2\omega^{(0)}_{I,322}$ is the superradiance rate of the Kerr $|322\rangle$ state, and the corrections from the cloud, $\gamma_1$ and $\gamma_2$, follow from the imaginary parts of $\delta\omega^{(2)}_{322}$ and $\delta\omega^{(4)}_{322}$. For $\alpha\ll 1$, $\gamma_1$ comes mainly from the GW transition $322\to211$, which damps the $|322\rangle$ occupation at a rate $\propto N_{211}$.

The term quadratic in the cloud mass, $\gamma_2(\ms N_{211}/\mbh)^2$, is of fourth order in $\epsilon$. It is dominated by the gravitational scattering $|211\rangle\times|211\rangle\to|322\rangle\times|{\rm BH}\rangle$. We now describe this process and compute $\gamma_2$.

\begin{figure}[t]
 \centering\includegraphics[width=\columnwidth]{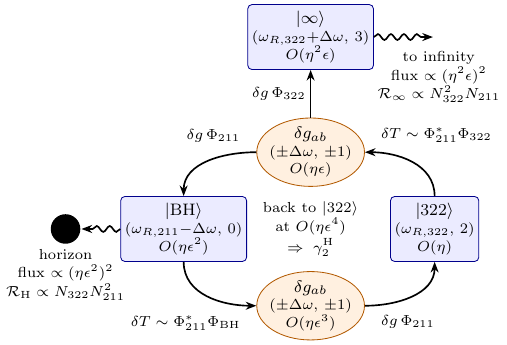}
 \caption{How gravity links the $|322\rangle$ mode, the $|211\rangle$ cloud, and the two scalar waves of Eq.~\eqref{eq:channels}. Boxes are scalar fields and ellipses are the metric perturbation $\delta g_{ab}$. Each gives the real part of the frequency, the azimuthal number, and the order of its field. Arrow labels are the source terms in the Einstein and Klein--Gordon equations. The horizon flux of the $|{\rm BH}\rangle$ wave defines $\gamma_2^{\rm H}$.}
 \label{fig:mode-conversion}
\end{figure}

\noindent{\bf{\em Gravitational mode conversion.--}}
The cloud self-gravity couples $|211\rangle$ and $|322\rangle$ through two dissipative scattering channels,
\begin{equation}
 \begin{aligned}
 &\text{absorption:}\quad &|211\rangle\times|211\rangle&\to|322\rangle\times|{\rm BH}\rangle,\\
 &\text{emission:}\quad &|322\rangle\times|322\rangle&\to|211\rangle\times|\infty\rangle ,
 \end{aligned}
 \label{eq:channels}
\end{equation}
Both channels proceed through the metric perturbation $\delta g_{ab}$, and the whole process is shown in Figure~\ref{fig:mode-conversion}. The process starts from a $|322\rangle$ perturbation $\Phi_{322}\equiv\eta\,\delta\Phi^{(0)}=O(\eta)$ on the $|211\rangle$ cloud $\Phi_{211}\equiv\Phi=O(\epsilon)$. The perturbed stress-energy tensor $\delta T_{ab}\sim\Phi_{211}^*\Phi_{322}+{\rm c.c.}$ sources $\delta g_{ab}=O(\eta\epsilon)$. Since $\Phi_{211}\propto e^{-i\omega_{R,211}t+i\varphi}$ and $\Phi_{322}\propto e^{-i\omega_{R,322}t+2i\varphi}$, $\delta g_{ab}$ has frequency and azimuthal number $\pm(\Delta\omega,1)$, where $\Delta\omega=\omega_{R,322}-\omega_{R,211}$. The metric perturbation $\delta g_{ab}$ then enters the Klein--Gordon equation, where the products $\delta g_{ab}\Phi_{211}=O(\eta\epsilon^2)$ and $\delta g_{ab}\Phi_{322}=O(\eta^2\epsilon)$ source the scalar field at these two orders. The $(-\Delta\omega,-1)$ component of $\delta g_{ab}$ lowers the frequency of $\Phi_{211}$ by $\Delta\omega$ and its azimuthal number by one. The result is the $|{\rm BH}\rangle$ wave at $O(\eta\epsilon^2)$, with $m=0$ and frequency $2\omega_{R,211}-\omega_{R,322}$ below $\ms$, so it cannot escape and falls through the horizon. The $(+\Delta\omega,+1)$ component of $\delta g_{ab}$ raises the frequency of $\Phi_{322}$ by $\Delta\omega$ and its azimuthal number by one. The result is the $|\infty\rangle$ wave at $O(\eta^2\epsilon)$, with $m=3$ and frequency $2\omega_{R,322}-\omega_{R,211}$ above $\ms$, so it escapes to infinity. The fluxes of the two waves are $O(\eta^2\epsilon^4)$ into the horizon and $O(\eta^4\epsilon^2)$ to infinity. Since $N_{211}\propto\epsilon^2$ and $N_{322}\propto\eta^2$ at leading order, these fluxes give the rates
\begin{equation}
 \mathcal R_{\rm H}=\gamma_2^{\rm H}\Bigl(\frac{\ms N_{211}}{\mbh}\Bigr)^2N_{322},\qquad
 \mathcal R_\infty=C\,\frac{N_{211}N_{322}^2}{\mbh^4}.
 \label{eq:rates}
\end{equation}
The absorption rate $\mathcal R_{\rm H}$ is linear in $N_{322}$ and therefore changes the growth rate of $|322\rangle$. The emission rate $\mathcal R_\infty$ is quadratic in $N_{322}$ and does not enter $\gamma_2$. We therefore split $\gamma_2=\gamma_2^{\rm H}-\delta\gamma_2$, where
$\gamma_2^{\rm H}$ denotes the absorptive part of $\gamma_2$, and
$\delta\gamma_2$ collects all other contributions at this order. As we
show in Supplemental Material S2, $\delta\gamma_2$ is suppressed by a
factor of $\alpha^4$ relative to $\gamma_2$ for $\alpha\ll1$, so that
$\gamma_2\simeq\gamma_2^{\rm H}$. We compute $\gamma_2^{\rm H}$ from the horizon flux of the $|{\rm BH}\rangle$ wave, the $m=0$ part of $\delta\Phi^{(2)}$. Supplemental Material S1 and S2 give the details. Relative to the leading-order Newtonian result of Foster and Rodd~\cite{Foster:2026nuz} ($\gamma_2^{\rm LO}=b_0\ms\alpha^5$, $b_0\simeq 0.134$), relativistic effects enhance $\gamma_2^{\rm H}$ by a factor of $1.9$ at $\alpha=0.1$ and $3.2$ at $\alpha=0.2$.

\noindent{\bf{\em Cloud evolution and GW emission.--}}
Self-gravity thus redistributes bosons among the levels and, consequently, shifts the GW power across emission lines. To follow this redistribution, we evolve the occupation numbers of the dominant levels together with the BH mass $\mbh$ and spin $\chi$. The evolution includes superradiance, gravitational scattering, and GW emission by boson annihilation and by transitions between levels. Each level obeys
\begin{equation}
 \dot N_i=2\omega_{I,i}(\mbh,\chi)\,N_i+\sum_p\Delta N_i^{(p)}\,\mathcal R_p,
 \label{eq:evol}
\end{equation}
where $2\omega_{I,i}$ is the superradiance rate of level $i$ at the instantaneous $\mbh$ and $\chi$, the sum runs over the scattering channels and the GW emission processes, $\Delta N_i^{(p)}$ is the change of $N_i$ in one process $p$, and $\mathcal R_p$ is its rate. For the two channels of Eq.~\eqref{eq:channels}, $(\Delta N_{211},\Delta N_{322})$ is $(-2,+1)$ for absorption and $(+1,-2)$ for emission, with the rates $\mathcal R_{\rm H}$ and $\mathcal R_\infty$ of Eq.~\eqref{eq:rates}. The BH mass and angular momentum change as $\dot \mbh=-\sum_i\omega_{R,i}\,2\omega_{I,i}N_i+\dot E_{\rm H}$ and $\dot J=-\sum_i m_i\,2\omega_{I,i}N_i+\dot J_{\rm H}$, where $\dot E_{\rm H}$ and $\dot J_{\rm H}$ are the energy and angular momentum of the scalar waves absorbed by the horizon.
For comparison, we also compute the standard superradiant evolution without self-gravity using the same levels and initial state~\cite{Arvanitaki:2014wva,Ficarra:2018rfu,Guo:2022mpr}.

Self-gravity also shifts the real frequency of each level. The frequency shift of $|211\rangle$ due to its own cloud has been computed relativistically~\cite{Cannizzaro:2023jle,May:2024npn}. The shift of $|322\rangle$ in the $|211\rangle$ cloud is the real part of $\epsilon^2\delta\omega^{(2)}_{322}$ in Eq.~\eqref{eq:pert}. In general, the shift is linear in the masses of all levels,
\begin{equation}
 \delta\omega_{R,i}=\sum_jK_{ij}\,\frac{\ms N_j}{\mbh},
 \label{eq:shift}
\end{equation}
where $\ms N_j$ is the rest mass in level $j$ and $K_{ij}$ is the shift of level $i$ per unit $\ms N_j/\mbh$ at fixed BH mass and spin. In Eq.~\eqref{eq:evol}, these shifts enter the superradiance rate via $2\omega_{I,i}\propto m_i\Omega_{\rm H}-(\omega_{R,i}+\delta\omega_{R,i})$. Because the cloud radiates GWs at the level frequencies and their differences, the shifts also move the GW lines: annihilation in level $i$ radiates at $2(\omega_{R,i}+\delta\omega_{R,i})$, and the transition $i\to j$ at $\omega_{R,i}+\delta\omega_{R,i}-\omega_{R,j}-\delta\omega_{R,j}$.

\begin{figure}[t]
 \centering
 \includegraphics[width=\columnwidth]{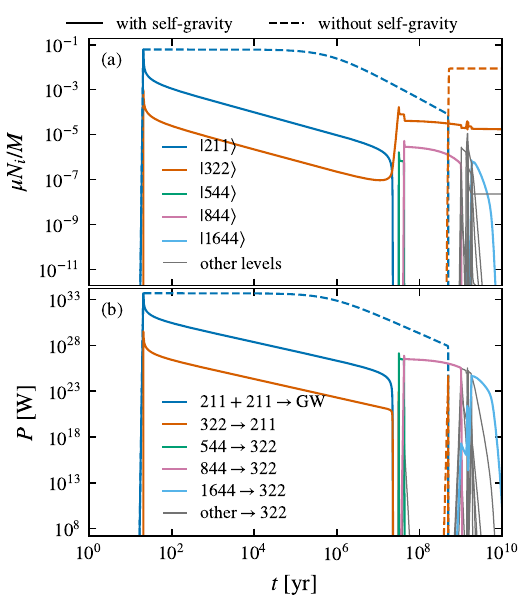}
 \caption{Evolution at $\alpha_0=0.1$ with (solid) and without (dashed) self-gravity: (a) occupations $\ms N_i/\mbh$ and (b) GW line powers. Grey lines show the levels and transitions not in the legends.}
 \label{fig:evolution}
\end{figure}

Superradiance first builds the $|211\rangle$ cloud. Absorption, with $(\Delta N_{211},\Delta N_{322})=(-2,+1)$ and $\mathcal R_{\rm H}\propto N_{211}^2N_{322}$, then boosts the extra growth of $|322\rangle$. Because emission ($\mathcal R_\infty\propto N_{211} N_{322}^2$) scales quadratically with $N_{322}$, it eventually overtakes absorption as $|322\rangle$ fills and stops its growth. The same absorption process with two $|322\rangle$ bosons, $|322\rangle\times|322\rangle\to|n44\rangle\times|{\rm BH}\rangle$, then feeds the $|n44\rangle$ levels, and their GW transitions $n44\to322$ return bosons to $|322\rangle$. We compute the dominant channels as above and take the others from the AxionSR code~\cite{Boskovic:2026boh}. We evolve 33 dominant levels~\cite{Foster:2026nuz} from $\mbh_0=10M_\odot$, $\chi_0=0.9$, and $\alpha_0=0.1$ or $0.2$, and Figure~\ref{fig:evolution} shows the resulting occupations and line powers for $\alpha_0=0.1$.
The $|211\rangle$ level loses half of its bosons within $0.25$ yr of its peak ($6.5\times10^5$ yr without self-gravity), $|322\rangle$ peaks at $\ms N_{322}/\mbh=9.8\times10^{-4}$ ($8.7\times10^{-3}$), $|544\rangle$ and $|844\rangle$ rise at $32$ and $42$ Myr, and $|1644\rangle$ rises at $1.8$ Gyr.

Self-gravity changes the GW lines through these occupations. The $211$ annihilation power scales as $N_{211}^2$, so a faster depletion of $|211\rangle$ weakens and shortens this line ($5\times10^5$ times less energy). The $322\to211$ and $n44\to322$ powers scale as $N_{322}N_{211}$ and $N_{n44}N_{322}$, so they depend on how long both levels of each pair stay populated together (the $322\to211$ line is $10^5$ times brighter at its peak but radiates $56$ times less energy).

\begin{figure*}[t]
 \centering\includegraphics[width=.90\textwidth]{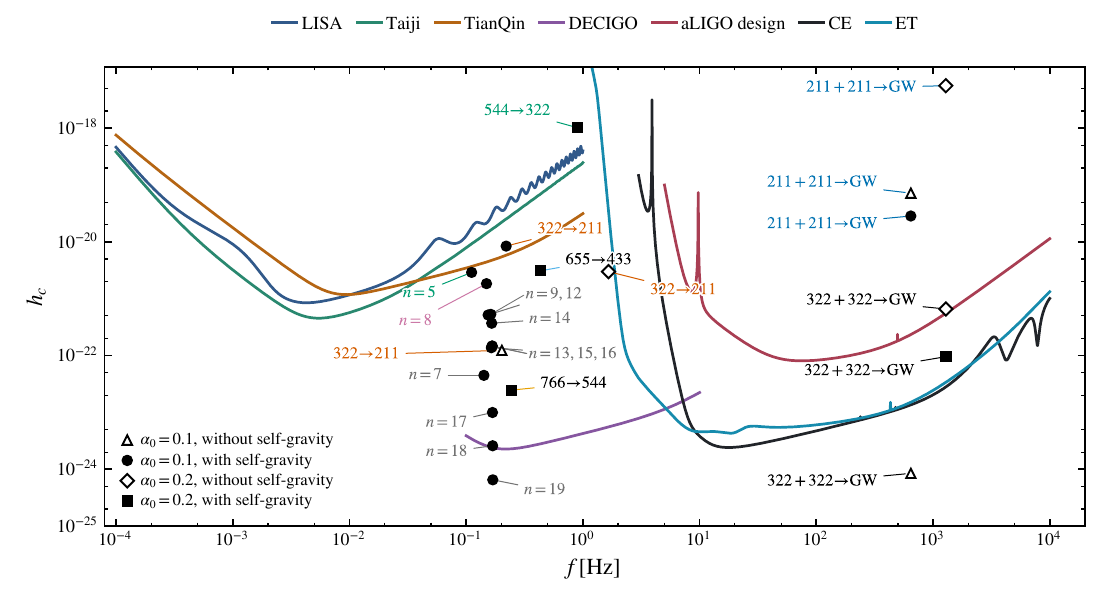}
 \caption{Four-year characteristic strain at 10 kpc for an initial BH mass of $10M_\odot$. At $\alpha_0=0.1$, open triangles and filled circles denote the models without and with self-gravity, respectively. At $\alpha_0=0.2$, open diamonds and filled squares denote the models without and with self-gravity, respectively, over the $10$ Gyr covered by the simulated cloud evolution. Labels $n$ denote $n44\to322$. For each model we include clouds up to $10$ Gyr and only show the channels that individually yield more than one expected event at $\rho\ge5$ in a four-year observation. The curves represent the noise at design sensitivity.}
 \label{fig:spectrum}
\end{figure*}

\noindent{\bf{\em GW spectroscopy.--}}
Annihilation from level $i$ radiates at $f_i\simeq{\omega_{R,i}}/\pi$ with power proportional to $N_i^2$. A transition $i\to j$ radiates at $f_{ij}\simeq ({\omega_{R,i}}-{\omega_{R,j}})/(2\pi)$ with power proportional to $N_iN_j$. To identify the strongest lines, we first combine the evolved occupations with weak-field annihilation rates and leading post-Newtonian radiation multipoles~\cite{Foster:2026nuz,Yoshino:2013ofa,Baryakhtar:2020gao}. At $\alpha_0=0.1$, $211$ annihilation gives the strongest ground-based signal, while the $322\to211$ and $n44\to322$ transitions give the strongest signals in DECIGO. 

We take the powers at infinity for $211$ and $322$ annihilation and the $322\to211$ transition directly from the evolution, which uses native Kerr radiation coefficients for these lines. The other transitions use the Kerr or AxionSR coefficients adopted in the evolution, including the AxionSR coefficient for $844\to322$ at both couplings~\cite{Boskovic:2026boh}. For the cloud with self-gravity at $\alpha_0=0.2$, we use powers and frequencies sampled every $1.0$ Myr from $2.0$ to $10$ Gyr to resolve the late-time oscillations.

\begin{figure*}[t]
 \centering\includegraphics[width=.90\textwidth]{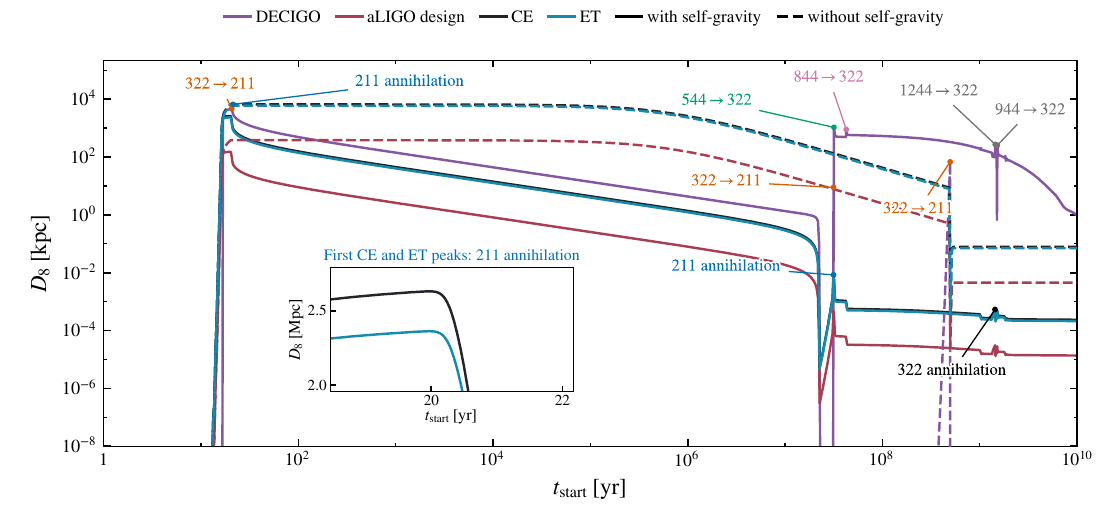}
 \caption{Distance at which the GWs give a four-year SNR of $\rho=8$, plotted against the age at the start of the observation for an initial BH mass of $10M_\odot$ and $\alpha_0=0.1$. Solid and dashed curves show the models with and without self-gravity, respectively. The total SNR combines all included lines within the same four-year observing window.}
 \label{fig:reach}
\end{figure*}

For a line of power $P_k$ at infinity, let $h_k^2=\langle h_{+,k}^2+h_{\times,k}^2\rangle$ denote the GW strain averaged over source orientations. At distance $D$, the strain and the signal-to-noise ratio (SNR) for an observation starting at cloud age $t_0$ are given by
\begin{align}
 h_k(t)&=\left( \frac{P_k(t)}{\pi^2D^2 f_k^2(t)}\right)^{1/2},\nonumber\\
 \rho^2(t_0)&=2\sum_k\int_{t_0}^{t_0+T}
       \frac{h_k^2(t)}{S_{\rm eff}[f_k(t)]}\,dt.
 \label{eq:reach}
\end{align}
Here $S_{\rm eff}$ is the one-sided strain noise power spectral density (PSD). We take an observation time of $T=4$ yr. Figure~\ref{fig:spectrum} shows the characteristic strain $h_{c,k}=(2\int f_k h_k^2\,dt)^{1/2}$. Figure~\ref{fig:reach} combines all the lines in the same observing window and gives the distance $D_8(t_0)=D\,\rho(t_0,D)/8$ at which their total SNR is $\rho=8$.

For the aLIGO interferometer we use $S_{\rm eff}=5S_{\rm inst}$~\cite{LIGOScientific:2014pky}. For one DECIGO triangle we use the PSD of \cite{Yagi:2011wg} over $0.1$--$10$ Hz and we adopt the design sensitivities for CE~\cite{Reitze:2019iox} and ET~\cite{Branchesi:2023mws}. For LISA, Taiji and TianQin we use two-channel sensitivities over $10^{-4}$--$1$ Hz, including the four-year Galactic foreground~\cite{Robson:2018ifk,Liu:2023qap,Huang:2020rjf}.

The $322\to211$  transition produces the strongest signal in the DECIGO band at $\alpha_0=0.1$. With self-gravity, it peaks near age $21$ yr at $0.22$ Hz, with a four-year SNR of $3.7\times10^3$ at 10 kpc. Without self-gravity, only the $322\to211$ transition exceeds $\rho=8$ in DECIGO at 10 kpc. Its peak four-year SNR is $54$. DECIGO can observe the cloud at $\rho=8$ out to $4.6$ Mpc with self-gravity. As a comparison, DECIGO can only detect such a source at $68$ kpc in the absence of self-gravity. 
For the $211$ annihilation line, it radiates at $650$ Hz and its four-year strain at 10 kpc is $2.9\times10^{-20}$ with self-gravity and $7.2\times10^{-20}$ without self-gravity. The $211$ annihilation line is above aLIGO, CE, and ET sensitivities with or without self-gravity correction.
However, the lifetime of this channel is largely suppressed after the self-gravity correction. The dashed lines in Figure~\ref{fig:reach} show that the $211$ annihilation channel is supposed to be the dominant mode in the absence of self-gravity. With self-gravity, the maximum distance at $\rho=8$ is greater for $322\to211$ in DECIGO than for $211$ annihilation in aLIGO, CE or ET. 

Both bands can therefore see the same young cloud. In the four years starting at age $19$ yr, $322\to211$  gives $\rho=3.5\times10^3$ in DECIGO and $211$ annihilation gives $\rho=1.2\times10^2$ in aLIGO at 10 kpc. Requiring each line to exceed $\rho=8$ gives a maximum distance of $0.15$ Mpc with self-gravity and $0.53$ kpc without self-gravity.
At later ages, the $544\to322$, $744\to322$ and $844\to322$ transitions peak at $0.11$, $0.14$ and $0.15$ Hz, respectively. At 10 kpc all three transitions exceed $\rho=8$ in DECIGO during a four-year observation. Near $42$ Myr, the window that maximizes the SNR of the weakest member of the triplet gives SNRs of $68$, $17$ and $7.1\times10^2$, respectively. At 10 kpc the maximum SNR of the 40 modeled lines with self-gravity is $0.20$ for LISA, $0.33$ for Taiji and $1.4$ for TianQin. These results indicate that self-gravity largely suppresses the annihilation channel and enhances the transition channels. At $\alpha_0=0.2$ and 10 kpc, $211$ annihilation peaks at $1.3$ kHz with $h_c=1.5\times10^{-19}$, and the strongest DECIGO line, $544\to322$, peaks at $0.90$ Hz with $h_c=1.0\times10^{-18}$ and $\rho=2.5\times10^5$.

To estimate how many distinct Galactic sources each detector could observe, we convert the star formation rate to a BH birth rate using the stellar initial mass function and remnant masses of Ref.~\cite{Yuan:2025fde}. We apply the trajectory with self-gravity to remnants born within $0.5M_\odot$ of $10M_\odot$ at $Z=0.8Z_\odot$. We adopt a uniform distribution of birth spin in $[0,1]$ and select spins $\chi_0\ge0.8$. At each cloud age we weight the selected birth rate by the fraction of the Galactic disk~\cite{McMillan:2016jtx} within which the lines give a total SNR of $\rho\ge5$. We adopt a star formation rate of $8.0\,M_\odot\,{\rm yr}^{-1}$~\cite{Siegert:2023dqg}. This benchmark assumes isolated evolution, a fixed remnant yield and disk profile, and individually distinguishable signals. We expect $0.36$ sources for aLIGO, $83$ for CE, $68$ for ET and $3.5\times10^6$ for DECIGO at $\alpha_0=0.1$. At $\alpha_0=0.2$, the corresponding counts are $2.8\times10^{-3}$, $2.1$, $1.2$ and $2.5\times10^6$, respectively. With self-gravity, the expected counts for LISA, Taiji and TianQin are $(1.6,\,5.5,\,1.3\times10^2)$ at $\alpha_0=0.1$ and $(0.017,\,0.094,\,1.0\times10^1)$ at $\alpha_0=0.2$. The $1644\to322$ and $766\to544$ lines alone give $2.1\times10^6$ and $2.2\times10^6$ DECIGO sources at $\alpha_0=0.1$ and $0.2$, respectively. At $\alpha_0=0.2$, the $544\to322$ line alone gives $2.6\times10^5$ DECIGO sources. Clouds aged $9.0$--$10$ Gyr account for $20\%$ of the total DECIGO count at this coupling.

\noindent{\bf{\em Discussion and conclusion.--}}
The annihilation channel and the transition channels appear at different epochs and emit GWs in different bands. Together, the two channels provide a multiband GW probe of ultralight dark matter. 
Current limits on the boson mass from continuous-wave searches for individual clouds~\cite{LIGOScientific:2021rnv} and from searches for a stochastic GW background~\cite{Tsukada:2018mbp,Yuan:2022bem} depend on the predicted annihilation signal without self-gravity. In this letter, we show that self-gravity suppresses the $|211\rangle$ channel and enhances decihertz transitions. This redistribution not only changes the signal duration time but also the strain of GWs. The cloud evolution under the self-gravity corrections also sets the initial conditions for resonant interactions with an inspiraling companion, which can shift the GW phase~\cite{Xu:2026cky,Xu:2026aic,Li:2026foy}.
For continuous GW searches, the cloud self-gravity correction also shifts the GW frequencies, indicating such a correction would affect the inferred boson mass. For instance, the $322\to211$  transition frequency is $8.4\%$ higher than its vacuum Kerr value at $\alpha_0=0.1$. Since the leading transition frequency scales as $\mbh^2\ms^3$, using the vacuum Kerr value in GW data analysis would lead to uncertainties of a few percent.
Extending these calculations across boson masses, BH masses and birth spins would yield a revised prediction for the stochastic GW background from the BH population~
\cite{Brito:2017wnc,Brito:2017zvb,Yuan:2021ebu}. Comparing this prediction with observational upper limits would give updated constraints on the boson mass~\cite{Tsukada:2018mbp,Yuan:2022bem}.

{\bf{Acknowledgments.--}}
During the preparation of this work, some of the authors used Claude, GPT, and Gemini to assist with language editing and computational tasks. All calculations and results were independently verified by the authors, who take full responsibility for the content of this manuscript. This work was partially
supported by NSFC (Grant No.1227516 and 1275074). 
Z.Z.\ acknowledges financial support from the MUR FIS2 Advanced Grant ET-NOW (CUP:~B53C25001080001) and the INFN TEONGRAV initiative. The code used in this work will be made publicly available in a revised version of this manuscript.

\bibliography{refs}

\clearpage
\onecolumngrid
\begin{center}
\textbf{\large Supplemental Material}
\end{center}
\setcounter{equation}{0}
\renewcommand{\theequation}{S\arabic{equation}}
\renewcommand{\theHequation}{S\arabic{equation}}
\setcounter{section}{0}
\setcounter{secnumdepth}{2}
\renewcommand{\thesection}{S\arabic{section}}
\renewcommand{\theHsection}{S\arabic{section}}
\setcounter{table}{0}
\renewcommand{\thetable}{S\Roman{table}}
\renewcommand{\theHtable}{S\Roman{table}}
\setcounter{figure}{0}
\renewcommand{\thefigure}{S\arabic{figure}}
\renewcommand{\theHfigure}{S\arabic{figure}}

\section{Fourth-order perturbation theory for the \texorpdfstring{$|322\rangle$}{|322>} mode}
\label{sec:s1}
As in the main text, we expand the synchronized solution~\cite{Herdeiro:2014goa} and the $|322\rangle$ mode on it in the cloud amplitude $\epsilon$ about the Kerr metric $g^{(0)}_{ab}$, on which the $|211\rangle$ state has the real frequency $\omega^{(0)}_{R,211}=\Omega_{\rm H}$, at fixed $\mu$ and Christodoulou mass $M$, and solve the metric equations in Lorenz gauge~\cite{Dolan:2023enf,Wardell:2024yoi}.
We normalize the amplitude by $\epsilon^2=\mu N_{211}/M$, so that $\epsilon^2$ is the rest mass of the cloud in units of $M$.
At $O(\eta\epsilon)$, the perturbed stress-energy tensor contains the cross terms $\Phi_{211}^*\Phi_{322}+{\rm c.c.}$ of the cloud and the mode.
With $\Phi_{211}\propto e^{-i\omega_{R,211}t+i\varphi}$ and $\Phi_{322}\propto e^{-i\omega_{322}t+2i\varphi}$, $\Phi_{211}^*\Phi_{322}$ oscillates as $e^{-i(\omega_{322}-\omega_{R,211})t+i\varphi}$, at the difference frequency and with $m=1$, and its complex conjugate has $m=-1$.
The stress-energy tensor is real, and so is the metric perturbation $\delta g^{(1)}_{ab}$ it sources. It splits into an $m=1$ part, sourced by $\Phi_{211}^*\Phi_{322}$, and its complex conjugate, with $m=-1$:
\begin{equation}
 \delta g^{(1)}_{ab}=\delta g^{(1)+}_{ab}+\delta g^{(1)-}_{ab},\qquad \delta g^{(1)-}_{ab}=\bigl(\delta g^{(1)+}_{ab}\bigr)^*,
 \label{eq:s2-pm}
\end{equation}
and $\delta g^{(1)}_{ab}$, ingoing at the horizon and outgoing at infinity, solves
\begin{equation}
 \mathcal E^{(0)}[\delta g^{(1)}]_{ab}=8\pi\,\delta T^{(1)}_{ab}.
 \label{eq:s2-g1}
\end{equation}
From here on, each field is written as a spatial profile times $e^{-i\omega t+im\varphi}$, and $\mathcal O^{(0)}$ and $\mathcal E^{(0)}$ denote $\Box_{g^{(0)}}-\mu^2$ and the linearized Einstein operator acting on the profile at fixed $\omega$ and $m$.
Each is evaluated at the Kerr frequency of the field it acts on: $\omega^{(0)}_{322}$ for $m=2$, $\omega^{(0)}_{322}-\omega^{(0)}_{R,211}$ for the $m=1$ metric, and $2\omega^{(0)}_{R,211}-\omega^{(0)*}_{322}$ for the $m=0$ wave.
The frequency shifts then enter through $\omega$-derivatives of the operators at fixed profiles, e.g.,
\begin{equation*}
 \mathcal O^{(0)}(\omega_{322})=\mathcal O^{(0)}(\omega^{(0)}_{322})+\epsilon^2\delta\omega^{(2)}_{322}\,\partial_\omega\mathcal O^{(0)}+\cdots .
\end{equation*}
At $O(\eta\epsilon^2)$, the Klein--Gordon equation reads
\begin{equation}
 \mathcal O^{(0)}\delta\Phi^{(2)}=-\bigl(\delta\mathcal O[h^{(2)}]+\delta\omega^{(2)}_{322}\,\partial_\omega\mathcal O^{(0)}\bigr)\delta\Phi^{(0)}-\delta\mathcal O[\delta g^{(1)}]\,\Phi^{(1)},
 \label{eq:s2-kg}
\end{equation}
where $\delta\mathcal O[h]$ is the first-order change of the Klein--Gordon operator $\mathcal O=\Box_g-\mu^2$ under $g_{ab}\to g_{ab}+h_{ab}$~\cite{Cannizzaro:2023jle}; in Lorenz gauge, $\nabla^a(h_{ab}-\tfrac12 g^{(0)}_{ab}h)=0$ with $h=g^{(0)ab}h_{ab}$, it is $\delta\mathcal O[h]\Phi=-h^{ab}\nabla_a\nabla_b\Phi$.
Since $\delta\mathcal O$ is linear,
\begin{equation}
 \delta\mathcal O[\delta g^{(1)}]\,\Phi^{(1)}=\delta\mathcal O[\delta g^{(1)+}]\,\Phi^{(1)}+\delta\mathcal O[\delta g^{(1)-}]\,\Phi^{(1)}.
 \label{eq:s2-split}
\end{equation}
Multiplying the factor of each metric part by the cloud factor $e^{-i\omega_{R,211}t+i\varphi}$ gives
\begin{equation}
\begin{aligned}
 \delta\mathcal O[\delta g^{(1)+}]\,\Phi^{(1)}&\propto e^{-i(\omega_{322}-\omega_{R,211})t+i\varphi}\,e^{-i\omega_{R,211}t+i\varphi}=e^{-i\omega_{322}t+2i\varphi},\\
 \delta\mathcal O[\delta g^{(1)-}]\,\Phi^{(1)}&\propto e^{i(\omega_{322}^*-\omega_{R,211})t-i\varphi}\,e^{-i\omega_{R,211}t+i\varphi}=e^{-i\omega_{\rm BH}t},
\end{aligned}
 \label{eq:s2-harm}
\end{equation}
where $\omega_{\rm BH}=2\omega_{R,211}-\omega_{322}^*$ is the frequency of the $|{\rm BH}\rangle$ wave of the main text; $\omega_{322}$ enters conjugated because $\delta g^{(1)-}_{ab}=(\delta g^{(1)+}_{ab})^*$ and $\omega_{322}$ is complex.
On the axisymmetric Kerr background, Eq.~\eqref{eq:s2-kg} therefore separates into an $m=2$ equation for $\delta\Phi^{(2)}_{322}$, the $O(\eta\epsilon^2)$ correction to the $|322\rangle$ mode $\delta\Phi^{(0)}$, and an $m=0$ equation for the wave $\delta\Phi^{(2)}_{\rm BH}$, which falls into the horizon:
\begin{equation}
\begin{aligned}
 \mathcal O^{(0)}\delta\Phi^{(2)}_{\rm BH}&=-\delta\mathcal O[\delta g^{(1)-}]\,\Phi^{(1)},\\
 \mathcal O^{(0)}\delta\Phi^{(2)}_{322}&=S^{(2)}_{322},\\
 S^{(2)}_{322}&=-\bigl(\delta\mathcal O[h^{(2)}]+\delta\omega^{(2)}_{322}\,\partial_\omega\mathcal O^{(0)}\bigr)\delta\Phi^{(0)}-\delta\mathcal O[\delta g^{(1)+}]\,\Phi^{(1)}.
\end{aligned}
\label{eq:s2-low}
\end{equation}
At $O(\eta\epsilon^3)$, the second-order fields source the metric perturbation $\delta g^{(3)}_{ab}$. The $m=1$ part of the source contains $\Phi^{(1)}\delta\Phi^{(2)*}_{\rm BH}$, so the $|{\rm BH}\rangle$ wave feeds back into the $m=2$ equation through $\delta g^{(3)+}_{ab}$. As in Eq.~\eqref{eq:s2-pm}, $\delta g^{(3)}_{ab}=\delta g^{(3)+}_{ab}+\delta g^{(3)-}_{ab}$ with $\delta g^{(3)-}_{ab}=(\delta g^{(3)+}_{ab})^*$; only the $m=1$ part enters the $m=2$ equation at $O(\eta\epsilon^4)$, and it obeys
\begin{equation}
\begin{aligned}
 \mathcal E^{(0)}[\delta g^{(3)+}]_{ab}
 &=8\pi\,\delta T^{(3)+}_{ab}-\delta^2G_{ab}[h^{(2)},\delta g^{(1)+}]\\
 &\quad-(\delta\omega^{(2)}_{322}-\delta\omega^{(2)}_{R,211})\,\partial_\omega\mathcal E^{(0)}[\delta g^{(1)+}]_{ab}.
\end{aligned}
 \label{eq:s2-g3}
\end{equation}
Here $\delta T^{(3)+}_{ab}$ is the $m=1$ part of $\delta T^{(3)}_{ab}$ in
\begin{equation*}
 \delta T_{ab}=\eta\,\bigl[\epsilon\,\delta T^{(1)}_{ab}+\epsilon^3\delta T^{(3)}_{ab}+\cdots\bigr],
\end{equation*}
and $\delta^2G_{ab}[h,h']$ is the second-order change of the Einstein tensor, bilinear in the two metric perturbations:
\begin{equation}
 \delta^2G_{ab}[h,h']=\frac{\partial^2}{\partial s\,\partial t}\,G_{ab}\bigl[g^{(0)}+s\,h+t\,h'\bigr]\Big|_{s=t=0}.
 \label{eq:s2-d2G}
\end{equation}
At $O(\eta\epsilon^4)$, the $m=2$ equation contains the frequency shift $\delta\omega^{(4)}_{322}$:
\begin{equation}
\begin{aligned}
 \mathcal O^{(0)}\delta\Phi^{(4)}_{322}&=S^{(4)}_{322},\\
 S^{(4)}_{322}&=-\delta\omega^{(4)}_{322}\,\partial_\omega\mathcal O^{(0)}\,\delta\Phi^{(0)}
 -\bigl(\delta\mathcal O[h^{(2)}]+\delta\omega^{(2)}_{322}\,\partial_\omega\mathcal O^{(0)}\bigr)\,\delta\Phi^{(2)}_{322}\\
 &\quad-\Bigl(\delta\mathcal O[h^{(4)}]+\tfrac12\delta^2\mathcal O[h^{(2)},h^{(2)}]
 +\delta\omega^{(2)}_{322}\,\partial_\omega\delta\mathcal O[h^{(2)}]
 +\tfrac12\bigl(\delta\omega^{(2)}_{322}\bigr)^2\partial_\omega^2\mathcal O^{(0)}\Bigr)\,\delta\Phi^{(0)}\\
 &\quad-\delta\mathcal O[\delta g^{(3)+}]\,\Phi^{(1)}-\delta\mathcal O[\delta g^{(1)+}]\,\Phi^{(3)}
 -\delta^2\mathcal O[h^{(2)},\delta g^{(1)+}]\,\Phi^{(1)}\\
 &\quad-\bigl(\delta\omega^{(2)}_{322}\partial_{\omega_{322}}+\delta\omega^{(2)}_{R,211}\partial_{\omega_{R,211}}\bigr)
 \bigl(\delta\mathcal O[\delta g^{(1)+}]\Phi^{(1)}\bigr).
\end{aligned}
\label{eq:s2-o4}
\end{equation}
Here $\delta^2\mathcal O[h,h']$ is the second-order change of $\mathcal O$, defined in the same way.
For $n=2,4$, $\delta\Phi^{(n)}_{322}$ obeys the boundary conditions of the $|322\rangle$ mode, ingoing at the horizon and decaying at infinity.
With these conditions $\mathcal O^{(0)}\delta\Phi^{(0)}=0$, so $\mathcal O^{(0)}$ has a zero mode, and $\mathcal O^{(0)}\delta\Phi^{(n)}_{322}=S^{(n)}_{322}$ has a solution only if its source satisfies a solvability condition.
We pair profiles bilinearly, without complex conjugation~\cite{Green:2022htq,Cannizzaro:2023jle},
\begin{equation}
 \langle\Psi,X\rangle=\int_{r_+}^\infty\!dr\int_0^\pi\!d\theta\;\Psi(r,\theta)\,X(r,\theta),
 \label{eq:s2-pair}
\end{equation}
where $r_+$ is the outer horizon radius, and define the left zero mode $\Psi$ by $\langle\Psi,\mathcal O^{(0)}X\rangle=0$ for every $X$ with these conditions.
Pairing $\Psi$ with $\mathcal O^{(0)}\delta\Phi^{(n)}_{322}=S^{(n)}_{322}$, the left side vanishes by the definition of $\Psi$, which gives the solvability condition
\begin{equation}
 \bigl\langle\Psi,S^{(n)}_{322}\bigr\rangle=0,\qquad n=2,4.
\label{eq:s2-solv}
\end{equation}
Since $S^{(n)}_{322}$ contains $-\delta\omega^{(n)}_{322}\,\partial_\omega\mathcal O^{(0)}\,\delta\Phi^{(0)}$, this condition fixes $\delta\omega^{(n)}_{322}$:
\begin{equation*}
 \delta\omega^{(n)}_{322}=\frac{\bigl\langle\Psi,\,S^{(n)}_{322}\big|_{\delta\omega^{(n)}_{322}=0}\bigr\rangle}{\bigl\langle\Psi,\,\partial_\omega\mathcal O^{(0)}\,\delta\Phi^{(0)}\bigr\rangle}.
\end{equation*}
With $\epsilon^2=\mu N_{211}/M$, the coefficients of the main text are $\gamma_1=2\operatorname{Im}\delta\omega^{(2)}_{322}$ and $\gamma_2=2\operatorname{Im}\delta\omega^{(4)}_{322}$.

\section{Flux balance and the complete coefficient}
\label{sec:s2}

The split $\gamma_2=\gamma_2^{\rm H}-\delta\gamma_2$ of the main text follows from a flux balance.
The synchronized background, with $\Phi\propto e^{-i\omega_{R,211}t+i\varphi}$ and an axisymmetric metric, is invariant under a rotation combined with a phase rotation of the scalar field~\cite{Herdeiro:2014goa},
\begin{equation}
 \varphi\to\varphi+\lambda,\qquad \Phi\to e^{-i\lambda}\,\Phi,
 \label{eq:s3-sym}
\end{equation}
for any real $\lambda$. The rotation generates the angular momentum $J$ of the scalar and metric perturbations and the phase rotation the scalar particle number $N$, so the perturbation carries the conserved charge $Q\equiv J-N$~\cite{Hollands:2012sf}.
A boson in the state $|n\ell m\rangle$, whose field depends on $\varphi$ as $e^{im\varphi}$, carries angular momentum $m$; the $|211\rangle$ cloud therefore has $J=N$ and carries no $Q$. Each $|322\rangle$ boson, with $m=2$, carries $Q=+1$, and each boson of the $m=0$ wave $|{\rm BH}\rangle$ carries $Q=-1$, so the scalar part of $Q$ counts $|322\rangle$ bosons minus $|{\rm BH}\rangle$ bosons.
The two channels of Eq.~(6) of the main text,
\begin{equation}
 \begin{aligned}
 &\text{absorption:}\quad &|211\rangle\times|211\rangle&\to|322\rangle\times|{\rm BH}\rangle,\\
 &\text{emission:}\quad &|322\rangle\times|322\rangle&\to|211\rangle\times|\infty\rangle,
 \end{aligned}
 \label{eq:s3-channels}
\end{equation}
both conserve $Q$: absorption turns $0+0$ into $(+1)+(-1)$, and emission turns $(+1)+(+1)$ into $0+(+2)$, since the $m=3$ wave $|\infty\rangle$ carries $Q=+2$.
The charge $Q$ is bilinear in the perturbation and its complex conjugate~\cite{Hollands:2012sf}. The full $m=2$ scalar component $\delta\Phi_{322}\propto e^{-i\omega_{322}t+2i\varphi}$, the full $m=0$ component $\delta\Phi_{\rm BH}\propto e^{-i\omega_{\rm BH}t}$ with $\operatorname{Im}\omega_{\rm BH}=\omega_{I,322}$, and the metric perturbation all grow as $e^{\omega_{I,322}t}$, so $dQ/dt=2\omega_{I,322}Q$.
Since $Q$ is conserved, its value outside the horizon changes only through the flux $F_{\rm H}$ of $Q$ into the horizon and the flux $F_\infty$ to infinity:
\begin{equation}
 \frac{dQ}{dt}=2\omega_{I,322}\,Q=-(F_{\rm H}+F_\infty).
 \label{eq:s3-balance}
\end{equation}
With $\epsilon^2=\mu N_{211}/M$, Eq.~(5) of the main text reads
\begin{equation}
 2\omega_{I,322}=2\omega^{(0)}_{I,322}+\gamma_1\epsilon^2+\gamma_2\epsilon^4+\cdots.
 \label{eq:s3-eq5}
\end{equation}
We expand $Q=Q_0+\epsilon^2Q_1+\epsilon^4Q_2+\cdots$ and the fluxes in the same way, $F=F_0+\epsilon^2F_1+\epsilon^4F_2+\cdots$. Because $\mathcal O^{(0)}\delta\Phi^{(0)}=0$, the amplitude is free up to a rescaling with real $c_2$ and $c_4$,
\begin{equation}
 \eta\to\bigl(1+c_2\epsilon^2+c_4\epsilon^4\bigr)\,\eta,
 \label{eq:s3-rescale}
\end{equation}
which rescales the whole perturbation, scalar and metric. $Q$ and all fluxes therefore change by $(1+c_2\epsilon^2+c_4\epsilon^4)^2$, which shifts
\begin{equation}
 Q_1\to Q_1+2c_2Q_0,\qquad
 Q_2\to Q_2+2c_2Q_1+\bigl(c_2^2+2c_4\bigr)Q_0,
 \label{eq:s3-Qshift}
\end{equation}
and we choose $c_2$ and $c_4$ so that $Q_1=Q_2=0$.
Order by order, Eq.~\eqref{eq:s3-balance} then gives
\begin{equation}
 2\omega^{(0)}_{I,322}Q_0=-(F_{{\rm H},0}+F_{\infty,0}),\qquad
 \gamma_1Q_0=-(F_{{\rm H},1}+F_{\infty,1}),\qquad
 \gamma_2Q_0=-(F_{{\rm H},2}+F_{\infty,2}).
 \label{eq:s3-orders}
\end{equation}
The $|{\rm BH}\rangle$ wave carries the part $F_{{\rm BH},2}$ of $F_{{\rm H},2}$. Each $|{\rm BH}\rangle$ boson that falls into the horizon leaves one $|322\rangle$ boson outside, so this part makes $|322\rangle$ grow and gives $\gamma_2^{\rm H}$, and the other fluxes give the remainder $\delta\gamma_2$. Dividing $\gamma_2Q_0=-(F_{{\rm H},2}+F_{\infty,2})$ by $Q_0$ and separating $F_{{\rm BH},2}$ from $F_{{\rm H},2}$ gives
\begin{equation}
 \gamma_2=-\frac{F_{{\rm H},2}+F_{\infty,2}}{Q_0}
 =\underbrace{-\frac{F_{{\rm BH},2}}{Q_0}}_{\gamma_2^{\rm H}}
 -\underbrace{\frac{F_{{\rm H},2}-F_{{\rm BH},2}+F_{\infty,2}}{Q_0}}_{\delta\gamma_2}.
 \label{eq:s3-split}
\end{equation}
The horizon flux has three parts, carried by the $|{\rm BH}\rangle$ wave, the $|322\rangle$ mode and the metric perturbation,
\begin{equation*}
 F_{{\rm H},2}=F_{{\rm BH},2}+F_{322,2}+F_{g,2},
\end{equation*}
so $\delta\gamma_2=(F_{322,2}+F_{g,2}+F_{\infty,2})/Q_0$. For $\alpha\ll 1$
\begin{equation*}
 \frac{F_{322,2}}{Q_0}\sim\alpha^8\gamma_2^{\rm LO},\qquad
 \frac{F_{g,2}}{Q_0}\sim\alpha^{10}\gamma_2^{\rm LO},\qquad
 \frac{F_{\infty,2}}{Q_0}\sim\alpha^4\gamma_2^{\rm LO}.
\end{equation*}
The flux to infinity $F_{\infty,2}$, the $\epsilon^4$ coefficient of the 322$\to$211 GW flux, therefore dominates, and a Newtonian calculation gives
\begin{equation*}
 \delta\gamma_2\simeq\frac{F_{\infty,2}}{Q_0}\simeq1.6\times10^{-4}\alpha^4\gamma_2^{\rm LO},\qquad \gamma_2^{\rm LO}\simeq0.134\,\mu\alpha^5,
\end{equation*}
where $\gamma_2^{\rm LO}$ is the leading Newtonian coefficient~\cite{Foster:2026nuz}. Hence $\gamma_2\simeq\gamma_2^{\rm H}$.
The complete numerical calculation on Kerr gives $\gamma_2/\gamma_2^{\rm LO}=1.91$, $3.20$ and $1.24$ at $\alpha=0.1$, $0.2$ and $0.3$.

For $\alpha\ll 1$, the leading-order relativistic correction to $\gamma_2^{\rm H}$ shows analytically where this enhancement comes from.
Since $\operatorname{Re}\omega_{\rm BH}<\mu$, the $|{\rm BH}\rangle$ wave is bound in the far region, like a superradiant level, and its horizon flux follows from matching the near-horizon and far-region solutions, as for the superradiant rates of Refs.~\cite{Detweiler:1980uk,Baumann:2019eav}.
At leading order this gives $\gamma_2^{\rm LO}$~\cite{Foster:2026nuz}.
At the next order, relativistic $1/r^2$ terms in the far region give the $\ell=0$ part of the $|{\rm BH}\rangle$ wave an extra factor
\begin{equation*}
 \Bigl(\frac{r}{r_{\rm B}}\Bigr)^{-6\alpha^2},\qquad M\ll r\ll r_{\rm B}=\frac{M}{\alpha^2},
\end{equation*}
where $r_{\rm B}$ is the Bohr radius. Near the black hole, $r\sim M$, the wave and the horizon flux are therefore larger than without these terms by
\begin{equation*}
 \Bigl(\frac{M}{r_{\rm B}}\Bigr)^{-6\alpha^2}=\alpha^{-12\alpha^2},\qquad
 \alpha^{-24\alpha^2}=e^{-24\alpha^2\ln\alpha}\simeq1-24\alpha^2\ln\alpha,
\end{equation*}
and with the other $\alpha^2$ terms
\begin{equation}
 \frac{\gamma_2^{\rm H}}{\gamma_2^{\rm LO}}\simeq1-24\alpha^2\ln\alpha+8.56\,\alpha^2+\cdots .
 \label{eq:s4-series}
\end{equation}
Both corrections are positive for $\alpha\ll1$, so the relativistic corrections increase $\gamma_2^{\rm H}$.
At $\alpha=0.1$, Eq.~\eqref{eq:s4-series} gives $\gamma_2^{\rm H}/\gamma_2^{\rm LO}\simeq1.64$, against the complete value $\gamma_2/\gamma_2^{\rm LO}=1.91$.

\end{document}